\documentclass[10pt, oneside, reqno]{article}
\usepackage[a4paper, total={6in, 8in}]{geometry}
 \usepackage{graphicx}
\usepackage{amsmath,amsthm,amsfonts,amscd,amssymb,mathrsfs}
\usepackage{authblk}
\usepackage{enumerate}
\usepackage[all]{xypic}
\usepackage{algorithm}
\usepackage{algpseudocode}
\usepackage{booktabs} 
\usepackage{bbm}
\usepackage{physics}
\usepackage{tikz}
\usepackage{multirow}
\usepackage{nicefrac}
\usepackage{xfrac}
\usepackage{makecell}
\usepackage{geometry}
\usepackage{array}
\usepackage{tabularx}
\usepackage{mathdots}
\usepackage{mathtools}

\usepackage{multicol}

\usepackage{hyperref}
\hypersetup{colorlinks=true,citecolor=blue, urlcolor= cyan}

\newcolumntype{C}{>{\centering\arraybackslash}m{6em}}

\theoremstyle{definition}

\newcommand{\Ex}[1]{\langle{#1}\rangle}

\usepackage[backend=biber,
style=numeric,
sorting=none
]{biblatex}

\AtEveryBibitem{
    \clearfield{urlyear}
    \clearfield{urlmonth}
}

\title{Empirical Demonstration of Quantum Contextuality on NISQ Computers}

\author[$\dagger, \ddagger$]{Colm Kelleher}
\author[$\dagger,\ddagger, \diamond$]{Frédéric Holweck}

\affil[$\dagger$]{Université Marie et Louis Pasteur, UTBM, CNRS, Laboratoire Interdisciplinaire Carnot de Bourgogne ICB UMR 6303, 90010 Belfort, France}
\affil[$\ddagger$]{Université Bourgogne Europe, CNRS, Laboratoire Interdisciplinaire Carnot de Bourgogne ICB UMR 6303, 21000 Dijon, France}
\affil[$\diamond$]{Department of Mathematics and Statistics, Auburn University, Auburn\nolinebreak[4] (AL), U.\,S.\,A.}

\date{May 2025}

\begin{document}
\maketitle

\begin{abstract}
We present definitive violations of non-contextual hidden variable bounds in the latest generation of IBM noisy intermediate-scale quantum computers (NISQ). These violations are based on known tests for contextuality such as the Rio Negro inequality and pseudo-telepathic Mermin games. These are the first violations of the classical Mermin game on IBM NISQ computers, and the largest such violations for the Rio Negro inequality. The use of finite geometries proves instrumental in the development of more effective tests, with larger geometries providing sizeable datasets from which multiple distinct experiments can be compared. 
\end{abstract}

\begin{multicols}{2}

\section{Introduction}
Quantum contextuality is a phenomenon limiting the kinds of hidden variable models describing the measurement outcomes of quantum mechanics. It was discovered by Bell \cite{bell_problem_1966} and independently by Kochen and Specker \cite{Kochen_specker} in the middle of the last century. It describes the impossibility of a deterministic hidden variable model to describe the measurement outcomes of quantum mechanics independently of the set of compatible measurements within which the original is considered. Since then it has been connected with nonlocality \cite{cabello_converting_2021}, quantum computing \cite{howard_quantum_2013}, error-correcting codes \cite{howard_contextuality_2014}, random access codes \cite{gatti_random_2023}, the quantum pigeonhole principle \cite{aharonov_quantum_2014} and even the quantum Cheshire Cat phenomenon \cite{hance_contextuality_2023}. See \cite{Budroni21} for an overview of the topic.

Experimental demonstrations of contextuality have been undertaken with varying success in both the laboratory setting \cite{kirchmair_state_independent_2009, xue_synchronous_2023, xu_experimental_2022, bartosik_2009_exp_contextuality, wu_experimental_2024} and on NISQ computers \cite{laghaout_demonstration_2022,holweck_testing_2021}. In this work we present definitive empirical evidence for the absence of non-contextual hidden variable models (NCHV) on the recently released IBM quantum backends with Heron R2 processors \cite{ibmq}. Including -- as far as the authors are aware -- the first victory of the Mermin game on modern day NISQ computers. This work presents updated results to the tests described in \cite{muller2025hexagons}, \cite{Kelleher_2_qubit_games} and \cite{kelleher_exploiting_2024}, with brief descriptions given below. 

\end{multicols}
\begin{figure}[h!]
\centering
\includegraphics[width=0.32\textwidth]{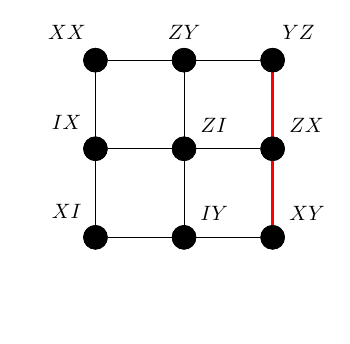}
\includegraphics[width=0.43\textwidth]{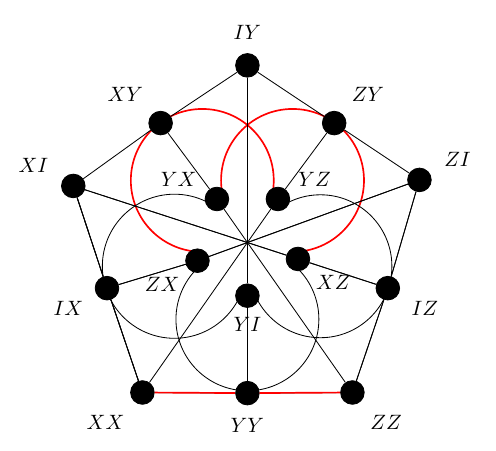}
\caption{The Peres-Mermin Magic Square (left) consisting of 9 operators (points) and 6 contexts (lines) of 3 operators each. The doily (right) with 15 points, 15 lines. Each context line gives a constraint on measurement outcome products, with black lines indicating $+1$ and red $-1$. The square has degree $d=1$, the doily $d=3$.}
\label{fig:mermin_square_doily}
\end{figure}
\begin{multicols}{2}

\subsection{Contextuality as Geometric Property}
The tests of contextuality in the current work arise naturally from the findings of Peres \cite{Peres90} and Mermin \cite{Mermin90, Mermin93} who gave a simple description of the phenomenon using just 9 Pauli spin operators. Consider the Peres-Mermin Magic Square in Fig. \ref{fig:mermin_square_doily} (left).

Each vertex is labelled by a 2-qubit Pauli spin operator with tensor product omitted, e.g. $IX = \sigma_{0}\otimes \sigma_{x}$. Triples of operators along a line pairwise commute and so such sets form \textit{contexts} of mutually-compatible operators. The product of any such set is $\pm \mathbbm{1}$, with the red line in Fig. \ref{fig:mermin_square_doily} (left) being the sole negative product, $(YZ)\cdot(ZX)\cdot(XY) = -\mathbbm{1}$. As the operators composing a context mutually commute, their product constraint must also be satisfied by their eigenvalues i.e. their simultaneous measurement outcomes. As such any deterministic hidden variable model assigning $\pm 1$ to each operator before measurement takes place must satisfy these context constraints. The square is ``magic" as the vertex set cannot be populated with non-contextual hidden variables from the set $\{+1, -1\}$ satisfying all constraints at once. This demonstrates the impossibility of such models, as any deterministic hidden variables must take into account the particular context considered for a given classical assignment. 

In the same fashion one can construct finite geometries of $N$-qubit operators, where lines are composed of triples of pairwise-commuting Pauli operators whose product is $\pm \mathbbm{1}$. The minimum number of unsatisfied line constraints for such a geometry is known as the \textit{degree} $d$ of contextuality, which for the Peres-Mermin magic square has $d=1$. As is clear in the study of such geometries, see for example \cite{henri_contextuality_2022, muller_multi_qubit_2022}, the contextuality of such sets of $N$-qubit operators is not dependent on the particular operator labelling of the vertices but rather the distribution of negative lines. The quantum operators furnish the negative line distribution, but all isomorphic quantum labelled geometries have identical degrees.

An important class of geometries are the symplectic polar spaces of rank $N$ over the field of characteristic 2, denoted $\mathcal{W}(2N-1,2)$, which encompass all nontrivial $N$-qubit operators as vertices and their commutation relations as lines. An example is the \textit{doily} $\mathcal{W}(3,2)$, depicted in Fig. \ref{fig:mermin_square_doily} (right) labelled by all nontrivial 2-qubit operators. All triples of pairwise commuting 2-qubit operators with product $\pm \mathbbm{1}$ are given by the set of 15 lines in the doily. See for example \cite{muller_multi_qubit_2022} for background.

Other contextual geometries worth examining are \textit{elliptic} and \textit{hyperbolic quadrics}, which have many interesting features as \textit{hyperplanes} of $\mathcal{W}(2N-1,2)$. Each quadric can be concisely described via a central point $p \in \mathcal{W}(2N-1,2)$. A quadric is the set of points $q \in \mathcal{W}(2N-1,2)$ that are either commuting with $p$ and \textit{symmetric} (an even number of $Y$s in its $N$-qubit operator) or anti-commuting with $p$ and \textit{skew-symmetric} (an odd number of $Y$s). Quadrics defined by a symmetric $p$ (including the trivial identity) are called \textit{hyperbolic} and denoted $H_{p}$, those defined by a skew $p$ are \textit{elliptic}, $E_{p}$. For any three points $q_{1}, q_{2}, q_{3}$ sharing a line in $\mathcal{W}(5,2)$ and contained in a quadric, the line is also considered a line of the quadric. Thus quadrics are sets of points described above plus all lines wholly incident with the sets. For $N=2$, hyperbolic quadrics are those subgeometries of the doily isomorphic to the Mermin square, of which there are 10, with that in Fig. \ref{fig:mermin_square_doily} being $H_{IX}$. Elliptic quadrics in the $N=2$ case form sets of 5 pairwise non-collinear points, of which there are 6 in $\mathcal{W}(3,2)$.

\section{The Rio Negro Inequality}
Experimental tests to demonstrate contextuality are well known. The two of interest to us in this work are the Rio Negro inequality \cite{cabello_proposed_2010} and the famed Mermin game \cite{cabello2001bell} both introduced by Cabello. In this section we present results on the former, following the protocol in \cite{muller2025hexagons} which improves upon the methods in \cite{laghaout_demonstration_2022, holweck_testing_2021}.

The Rio Negro inequality is an evaluation of the expectation values $\Ex{\mathcal{C}_{i}}$, the product of operators along the $i$th context of a given contextual geometry with $L$ lines and degree $d$. For positive $\mathcal{C}$ (negative $\mathcal{C}'$) contexts one expects $\Ex{\mathcal{C}_{i}} = +1$,  $\Ex{\mathcal{C}_{i}} = -1$. As such, one defines 

\begin{equation}
\chi := \sum_{i}\Ex{\mathcal{C}_{i}} - \sum_{i}\Ex{\mathcal{C}'_{i}}
\end{equation}

and expects, based on a NCHV or purely quantum mechanical (QM) model

\begin{equation}\label{eq:rio_negro_inequality}
\chi \leq \begin{cases} L - 2d, \text{ NCHV} \\
L, \quad \quad \; \text{ QM}
\end{cases}
\end{equation}

Larger values of $d$ relative to $L$ provide more opportunistic tests for the Rio Negro inequality, and so the above is examined for a variety of geometries. Essential among them is $\mathcal{W}(5,2)$, the symplectic polar space describing the commutation relations of all nontrivial 3-qubit operators (see \cite{saniga2006multiple} as well as cited work of the authors for further detail). This geometry contains 63 points, each labelled by a 3-qubit operator excluding $III$, and 315 lines, 90 of which are negative. It is known that the degree $d$ of $\mathcal{W}(5,2)$ is 63, the minimal set of unsatisfiable lines forming a configuration known as the split Cayley hexagon of order 2 \cite{muller2025hexagons}. Contained within $\mathcal{W}(5,2)$ are 3360 square configurations, each labelled with 3-qubit operators but with the same degree $d=1$. Of the configurations isomorphic to the doily, there are 1344 embedded in $\mathcal{W}(5,2)$, each of degree $d=3$. Finally, there are 28 elliptic quadrics ($L = 45, d = 9$) and 36 hyperbolic quadrics ($L = 105, d=21$) contained in $\mathcal{W}(5,2)$.

\subsection{Results}
\begin{figure}[h!]
\centering
\includegraphics[width=\textwidth]{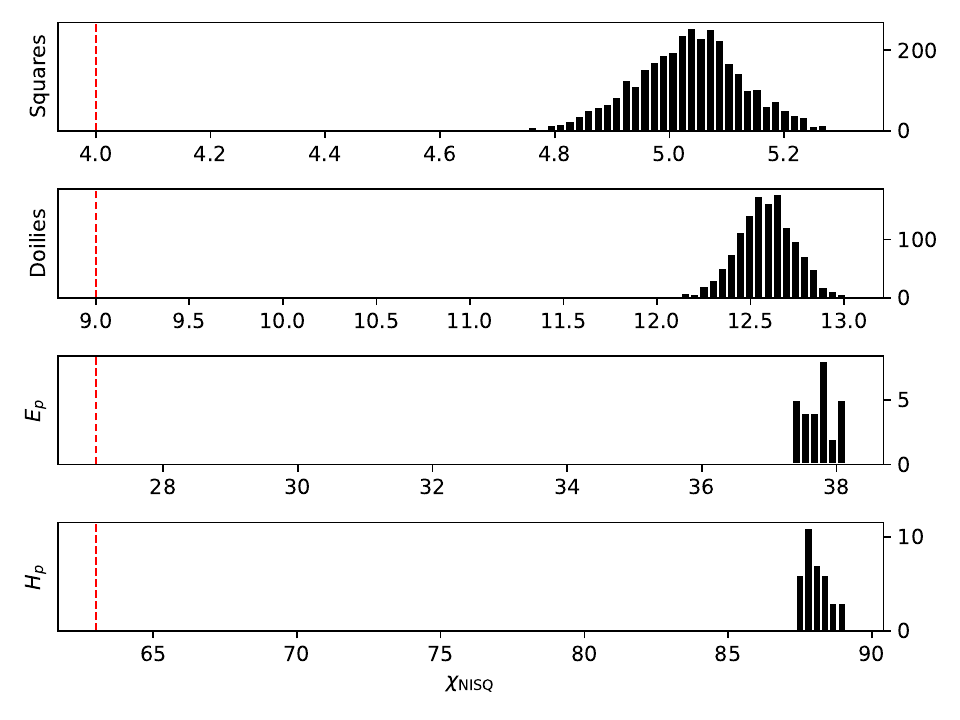}
\caption{Distributions for geometries squares, doilies, elliptic quadrics $E_{p}$ and hyperbolic quadrics $H_{p}$ vs. $\chi_{\text{NISQ}}$ values, from data computing $\Ex{\mathcal{C}}$ for each context in $\mathcal{W}(5,2)$. NCHV upper bounds shown in red, with each geometry clearly violating its bound.}
\label{fig:rio_negro_results}
\end{figure}

The procedure for testing the Rio Negro inequality on NISQ quantum circuits is outlined in \cite{muller2025hexagons}, where it was done on the IBM Eagle processors. Here we provide the results on the newer Heron R2 processors, on the full $\mathcal{W}(5,2)$ space of 3-qubit operators and subgeometries. The success rates are a clear improvement upon \cite{laghaout_demonstration_2022, holweck_testing_2021, muller2025hexagons} including a $10\%$ improvement on the second reference for $\mathcal{W}(5,2)$ and $H_{p}$ and $40\%$ on the latter reference for $E_{p}$. The results are given in Table \ref{tab:rio_negro_results} and Fig. \ref{fig:rio_negro_results}.


\end{multicols}
\begin{table}[h!]
\centering
\begin{tabular}{|c|cc|ccc|c|} \hline
    	$\mathcal{G}$ & $d$ & $L$ & $\chi_{\text{sim}}$ & $\chi_{\text{sim'}} $ & $\chi_{\text{NISQ}}$ & $L-2d$ \\ \hline
	$\mathcal{W}(5,2)$ & 63 & 315 & 315 & 269.5128 & $\boldsymbol{264.2206}$ & 189  \\
	Square & 1 & 6 & 6 & 5.1340 & $\boldsymbol{5.3076}$  & 4 \\
	Doily & 3 & 15 & 15 & 12.9230 & $\boldsymbol{13.0092}$ & 9 \\
	$E_{p}$ & 9 & 45 & 45 & 38.532 & $\boldsymbol{38.136}$  & 27 \\
	$H_{p}$ & 21 & 105 & 105 & 90.102 & $\boldsymbol{89.099}$  & 63 \\ \hline
    \end{tabular}\caption{Results for computing $\chi$ on various geometries on a noiseless simulator ($\chi_{\text{sim}}$), a simulator with noise model ($\chi_{\text{sim}'}$) and finally the Heron R2 backend \texttt{ibm\_kingston} ($\chi_{\text{NISQ}}$). In all cases each test was run with 10,000 shots. For subgeometries of $\mathcal{W}(5,2)$, best performing representative shown. In all cases the experimental result in bold violates its NCHV upper bound \eqref{eq:rio_negro_inequality}, indicating a lack of non-contextual hidden variables.}
\label{tab:rio_negro_results}
\end{table}

\vspace{2cm}

\begin{multicols}{2}

\section{Mermin-Like Games}
A second test of contextuality is the construction of no-communication, multi-player ``pseudotelepathy" games, where the players have a quantitative advantage when using quantum resources. In previous work \cite{Kelleher_2_qubit_games, kelleher_exploiting_2024} we constructed geometric variations of the Mermin game \cite{cabello2001bell} based on the square. In it, a referee chooses a pair of intersecting lines of the square and communicates one each to two spacelike separated players. In this scenario, known as the ``line-line" ($ll$) game, players must return triplets of values from the set $\{ +1, -1\}$ such that their product matches that of the line constraint. If the players' values match on their (unknown) intersection point, they win the game. It is clear that when playing with a classically assigned square (representing a deterministic NCHV model) they have $\omega_{ll}(S) = \frac{8}{9}$ success probability due to the unsatisfied constraint. But when playing with the quantum-labelled square Fig. \ref{fig:mermin_square_doily} (left) they can make measurements on a shared quantum state, ensuring agreement in all scenarios and winning with probability 1.

Similar games can be made for other geometries, with lines communicated to two players and success given if intersection values match. When playing on the doily for instance, NCHV success rate falls to $\omega_{ll}(D) = \frac{13}{15}$, while the quantum success remains at 1. Similarly for the 3-qubit elliptic quadric $E_{p}$ one has $\omega_{ll}(E) = \frac{13}{15}$. As $E_{p}$ has 5 lines incident with each point, a 4-player game can be constructed where the players win iff the product of their intersection values gives $+1$. This game has success probability of $\omega_{llll}(E)=\frac{11}{15}$. Finally, point-line versions were made for the square and doily, where one player is given a point and another an incident line and success occurs for matching values. In those cases, $\omega_{pl}(S) = \frac{17}{18}, \omega_{pl}(D)=\frac{14}{15}$. When implemented on a quantum circuit, any success rates above these bounds rules out NCHV models and proves the efficiency of quantum resources for these families of pseudo-telepathic games.

\begin{figure}[h!]
\centering
\includegraphics[width=0.72\textwidth]{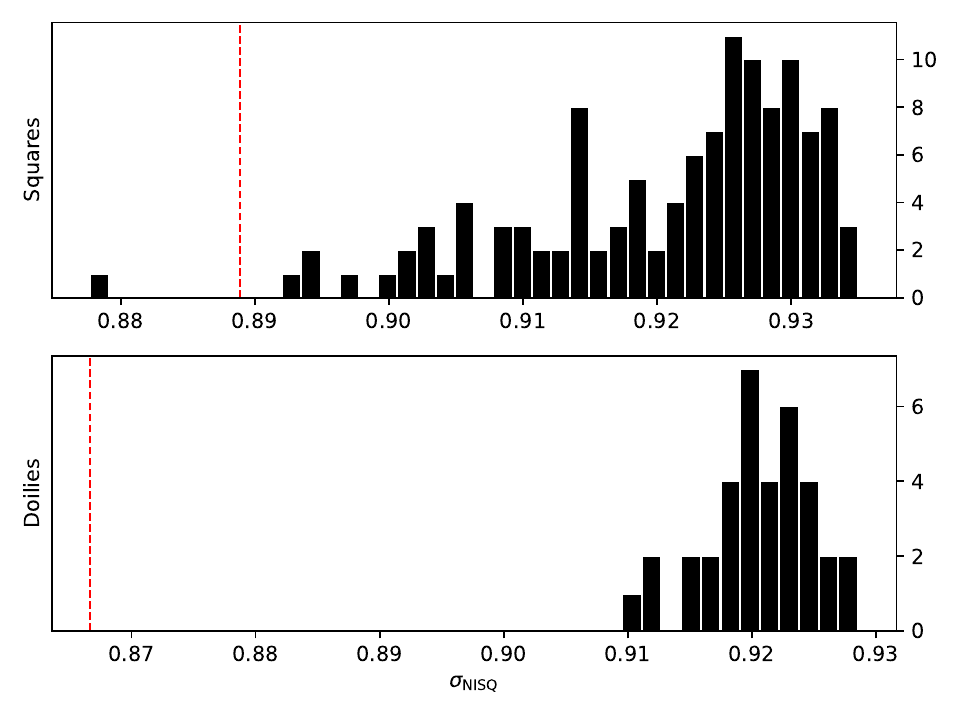}
\caption{Distributions for squares, doilies vs. $\sigma_{\text{NISQ}}$ values from $E_{YYY}$ $ll$ game. NCHV upper bounds shown in red, with all but one square violating its bound.}
\label{fig:doily_square_ll_results}
\end{figure}

Results of playing on the new Heron R2 processors are given in Table \ref{tab:mermin_game_results}. For the 2-player line-line ($ll$) games, all combinations of intersecting lines in $E_{YYY}$ were tested, and all doily and square subgeometry results were extracted and given in Fig. \ref{fig:doily_square_ll_results}.

\section{Conclusion}
In this paper we give updated results to a variety of quantum contextuality tests, laid out in \cite{muller2025hexagons}, \cite{Kelleher_2_qubit_games} and \cite{kelleher_exploiting_2024}. This is the first success of playing the Mermin and related games on NISQ computers, and improved violations of the Rio Negro inequality for larger, more complex geometries.

\end{multicols}
\begin{table}[h!]
\centering
\begin{tabular}{|c|c|c|ccc|c|} \hline
	\multirow{2}{*}{Game} & \multirow{2}{*}{Geometry} & \multirow{2}{*}{Backend} & \multicolumn{3}{c|}{Results} & \multirow{2}{*}{$\omega$}  \\ \cline{4-6}
      & & & $\sigma_{\text{sim}}$ & $\sigma_{\text{sim'}}$ & $\sigma_{\text{NISQ}}$ &  \\ \hline \hline
    \multirow{2}{*}{$pl$} & Square & \texttt{ibm\_marrakesh} & $1.0$ & $0.95628$ & $\boldsymbol{0.96527}$ & $0.9\overline{4}$ \\ 
    & Doily & \texttt{ibm\_marrakesh} & $1.0$ & $0.96160$ & $\boldsymbol{0.95945}$ & $0.9\overline{3}$ \\ \hline \hline
    \multirow{3}{*}{$ll$} & Square & \texttt{ibm\_marrakesh} & $1.0$ & $0.91722$ & $\boldsymbol{0.93498}$ & $0.\overline{88}$ \\ 
    & Doily & \texttt{ibm\_marrakesh} & $1.0$ & $0.92168$ & $\boldsymbol{0.92856}$ & $0.8\overline{6}$ \\
    & $E_{YYY}$ & \texttt{ibm\_marrakesh} & $1.0$ & $0.91988$ & $\boldsymbol{0.92067}$ & $0.8\overline{6}$ \\ \hline \hline
    $llll$ & $E_{YYY}$ & \texttt{ibm\_aachen} & $1.0$ & $0.78606$ & $\boldsymbol{0.84867}$ & $0.7\overline{3}$ \\ \hline
\end{tabular}
\caption{Results of playing the multi-player pseudotelepathy games on various geometries, on the indicated IBM Heron R2 backends with 10,000 shots per test. In $pl$ case, 2-qubit geometries Fig. \ref{fig:mermin_square_doily} were used. In $ll$ case, square and doily results were extracted from $E_{YYY}$ game with best performing results shown. In all games, the noiseless simulator $\sigma_{\text{sim}}$, noisy simulator $\sigma_{\text{sim}'}$ and indicated backend $\sigma_{\text{NISQ}}$ success rates violate the NCHV upper bounds $\omega$.}
\label{tab:mermin_game_results}
\end{table}

\begin{multicols}{2}

\section*{Acknowledgments}
This work is supported by the Graduate school EIPHI (contract ANR-17-EURE- 0002) through the project TACTICQ, the Ministry of Culture and Innovation. We acknowledge the use of the IBM Quantum Credits for this work. The views expressed are those of the authors and do not reflect the official policy or position of IBM or the IBM Quantum Experience team. The authors declare that they have no conflicts of interest to disclose, and would like to thank the developers of the open-source framework Qiskit. All code is available at \url{https://github.com/quantcert/quantcert.github.io}.

\printbibliography
\end{multicols}
\end{document}